\documentclass[10pt,journal,compsoc]{IEEEtran}
\ifCLASSOPTIONcompsoc
  \usepackage[nocompress]{cite}
\else
  \usepackage{cite}
\fi
\ifCLASSINFOpdf
\else
\fi
\usepackage{amsmath}
\DeclareMathOperator*{\argmax}{argmax} 
\usepackage{graphicx}
\usepackage{array}
\newcolumntype{C}[1]{>{\centering\arraybackslash}p{#1}}
\newcolumntype{L}[1]{>{\raggedright\arraybackslash}p{#1}}

\usepackage{xcolor}

\usepackage{array}
\usepackage{multirow}
\usepackage{amsfonts}
\begin{document}
%
\title{Echo Chambers and Segregation in Social Networks: Markov Bridge Models and Estimation}
%
%
%
%

\author{Rui~Luo,
        Buddhika~Nettasinghe,
        and~Vikram~Krishnamurthy,~\IEEEmembership{Fellow,~IEEE}
\IEEEcompsocitemizethanks{\IEEEcompsocthanksitem R. Luo is with the Sibley School of Mechanical and Aerospace Engineering, Cornell University, Ithaca, NY, 14850.\protect\\
E-mail: rl828@cornell.edu
\IEEEcompsocthanksitem B. Nettasinghe and V. Krishnamurthy are with the School of Electrical and Computer Engineering, Cornell University, Ithaca, NY, 14850.\protect\\
E-mail: \{dwn26, vikramk\}@cornell.edu
\IEEEcompsocthanksitem This research was supported by the U. S. Army Research
Office under grant W911NF-19-1-0365, and the National Science Foundation under grant 1714180.\protect\\}}

\IEEEtitleabstractindextext{%
\begin{abstract}
This paper deals with the modeling and estimation of the sociological phenomena called \emph{echo chambers} and \emph{segregation} in social networks. Specifically, we present a novel community-based graph model that represents the emergence of segregated echo chambers as a Markov bridge process. A Markov bridge is a one-dimensional Markov random field that facilitates modeling the formation and disassociation of communities at deterministic times which is important in social networks with known timed events. We justify the proposed model with six real world examples and examine its performance on a recent Twitter dataset. We provide model parameter estimation algorithm based on maximum likelihood and, a Bayesian filtering algorithm for recursively estimating the level of segregation using noisy samples obtained from the network. 
Numerical results indicate that the proposed filtering algorithm outperforms the conventional hidden Markov modeling in terms of the mean-squared error. The proposed filtering method is useful in computational social science where data-driven estimation of the level of segregation from noisy data is required.
\end{abstract}

}

\maketitle

\IEEEdisplaynontitleabstractindextext

%
\IEEEpeerreviewmaketitle

\IEEEraisesectionheading{\section{Introduction}\label{sec:introduction}}

%
%
%
%
\IEEEPARstart{O}NLINE Social networks (OSNs) lay the foundation for online community formation. Billions of users rely on OSNs to connect with friends, share information, and advertise products. Echo chambers, i.e., situations where one is exposed only to opinions that agree with their own, are an increasing concern for the usage of OSNs. According to the theory of preferential attachment or homophily \cite{mcpherson2001birds}, users tend to link with other users who share similar attributes (e.g.~opinions, interests, etc.). Further, social influence \cite{sasahara2019inevitability} also increases users' tendency of becoming more similar to somebody as a result of social interaction. These two factors lead to segregated and polarized clusters known as “echo chambers" on social networks. 

Echo chambers are studied on various social networks from different modelling perspectives. However, the evolution of echo chambers is characterised by certain temporal patterns in many cases, which is neglected by many proposed models. Having an anticipatory model of segregation in social networks allows us to incorporate the effects of periodic (i.e.,seasonal) events into the model. This enables the real time statistical inference as well as tasks such as offering incentives to reduce the effects of segregation (control strategies for preventing segregation and echo chambers). For example, one can imagine a control strategy which offers incentives to users at each time instant (subject to budget restrictions) to influence the link formation in order to hinder the segregation in social networks. 


Towards this end, the aim of this paper is to develop and analyze a model for the anticipatory nature of the segregation process in a social network~i.e.~one can assign probabilities to the event that the social network will be segregated at a certain fixed time instant.\footnote{A shorter version of this paper is under review for IEEE International Conference on Acoustics, Speech and Signal Processing~(2021).} 

\noindent{\bf Main Results:} 

(1) We present a dynamic network formation model that captures the dynamics of how a social network segregates into communities (and then integrates back again). The key idea behind our model is to represent the strength of the ties between communities (in terms of a graph clustering metric) as a Markov bridge process which is a special case of an anticipatory process.


(2) Based on the proposed Markov Bridge based segregation model, we propose a time-inhomogeneous Bayesian filter (called Hidden Markov bridge filter) for recursively estimating the state of the graph clustering metric. The Hidden Markov bridge filter uses only a few (compared to graph size) noisy samples from the social network at each time instant.

(3) We numerically compare the performance the proposed Hidden Markov bridge (HMB) filter with conventional Hidden Markov Model (HMM) filter in terms of mean-squared error. Our results show that the proposed method outperforms the traditional HMM filter. This shows that the Bayesian filter yields useful real time information about the dis-association and association of communities in a network.

(4) We evaluate the performance on a public available dataset \cite{chen2020election2020} which encompasses 7 million tweets related to the 2020 U.S. presidential election. Our results illustrate the proposed model and filter are useful in estimating Twitter users' state of political opinion polarization under real-world settings. 


\subsection{Why Markov Bridge Dynamics?}
In this subsection, we justify the usage of Markov bridge model in modeling social network phenomena based on six examples and a real world Twitter dataset.
The temporal dynamics of social networks give rise to states where the network is segregated into multiple communities at certain time instants and integrate back into a single community at other time instants. In statistical signal processing~(e.g.~in target tracking, etc.), dynamical processes with long-range dependencies are typically modeled as Markov bridge processes \cite{5947152,fanaswala2012detection,fanaswala2015spatiotemporal,fanaswala2011destination}. A Markov bridge process can be viewed as a special case of an anticipatory process in which a the distribution at a certain time instant is fixed. Below we discuss how the temporal dynamics of segregation in many  scenarios on social networks can be modeled using Markov bridge processes. 

\textit{Example 1. Schelling's Segregation Model: }
The segregation model developed by Schelling~\cite{schelling1971dynamic} is set in an $N\times N$ grid. Agents are split into two groups and occupy the spaces of the grid. Agents desire a fraction $B_{\textrm {a}}$ of their neighborhood to be from the same group. This model shows how segregated community and weak inter-community connectivity might arise even with a moderate individual preference $B_{\textrm {a}}$. The physical grid space can be generalized to a social network as a grid graph. In case of external stimulus such as elections, the value of $B_{\textrm {a}}$ or its distribution is anticipatory and can be modeled as a Markov bridge process.

\begin{figure}
	\centering
	\includegraphics[width=0.46\textwidth]{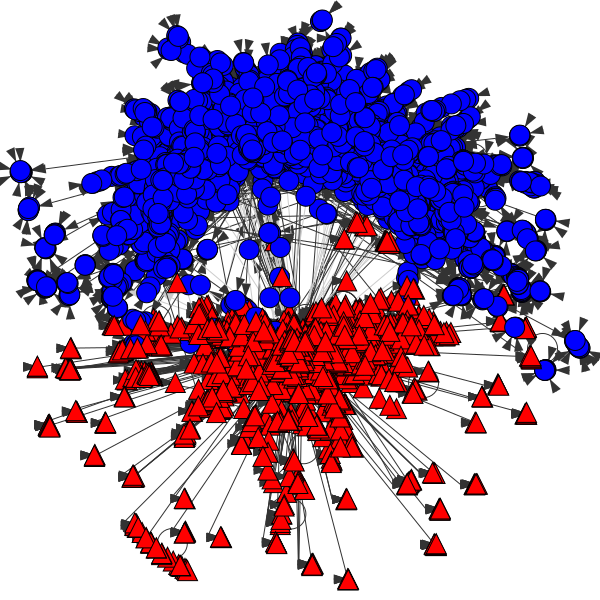}
	\caption{This figure indicates how Twitter users' political opinions are polarized into two communities (echo chambers) before the 2020 presidential election. Nodes represent Twitter users and edges represent retweets during the one-month period before election (Oct. 1 to Nov 1). Multiple snapshots illustrating the graph evolution during this period is shown in Fig. \ref{fig:fig4}. The graph is laid out using DrL (a force-directed graph layout) and the nodes are assigned different colors and shapes (blue circles and red triangles) according to the two communities detected by the Louvain method \cite{blondel2008fast}. Details can be found at section \ref{subsubsec:retweet network}.}
	\label{fig:fig1c}
\end{figure}

\textit{Example 2. Polarization of Political Opinion: }
\label{subsubsec: political}
OSNs (e.g. Twitter) users have different political leanings and tend to follow or retweet users of similar opinions. Such tendency can be moderate, i.e. users are open-minded to follow or hear someone from different ideology groups. However, during a politically polarization event, such as election or legislation, the tendency will become stronger and hinder users from connecting with others of different opinions, which results in highly segregated online echo chambers, as shown in Fig. \ref{fig:fig1c}. We can formulate the portion $x^{(t)}$ of interactions between users of different political leanings as a Markov bridge process. It is anticipatory to be at a low level during the polarization event.

\textit{Example 3. Spread of Coronavirus Fake News: }
On social media, people's perception of Coronavirus evolves during the out-break of this pandemic and leaves room for related fake news. Many factors prevent the spread of fake news, including scientific reports from reliable news sources, government releases, and users' tendency to share health and prevention messaging. We can model user's tendency $x^{(t)}$ to share Coronavirus fake news as a Markov bridge process. It will decrease with the elucidations from reliable sources and also bring down the probability that fake news echo chambers emerge. This example is related to the first one \ref{subsubsec: political} because it is found in \cite{jiang2020political} that partisanship correlates with sentiment toward government measures. Therefore, the evolution of Coronavirus fake news echo chambers is correlated with that of political ideology echo chambers.

\textit{Example 4. Active Level of Seasonal Sports League: }
Sports leagues typically have season and off-season. In different periods of a year, fans will have different involvement in the sports leagues on social media such as online sports forums. We can formulate the active level $x^{(t)}$ as a hidden state following a Markov bridge process. We can then use the data of posting, commenting, and time of stay on the forums as the observation. $x^{(t)}$ will evolve from a high level to a low level during the off-season, and will evolve back into a high level in the next sports season and lead to the segregated fan's community. Online merchants may take advantage of this information to maximize their advertisement coverage and return on investment.

\textit{Example 5. E-commerce Sequential Recommendation: }
Relevance and diversity usually act as two competing objectives in recommender systems, where the former causes growing concern that it might lead to the
self-reinforcing of user’s interests due to narrowed exposure of similar
items. The existence of echo chambers has been validated on user clicks, purchases, and browse logs from Alibaba Taobao in \cite{ge2020understanding}. To examine and quantify the echo chambers in recommender systems, we can use a measure $x^{(t)}$ to represent the similarity of recommended items during the interaction with users. For conventional recommender systems which narrow down the contents provided to users, $x^{(t)}$ can be modeled as a Markov bridge process which evolves from a low level to a high level. In this context, an extended recommendation framework can potentially avoid such echo chambers emergence by using collaborative filtering and sequential forecasting to recommend users items that they may find useful in the future, thus improving both user satisfaction and E-commerce platform's revenue.

\begin{figure}
	\centering
	\includegraphics[width=0.5\textwidth]{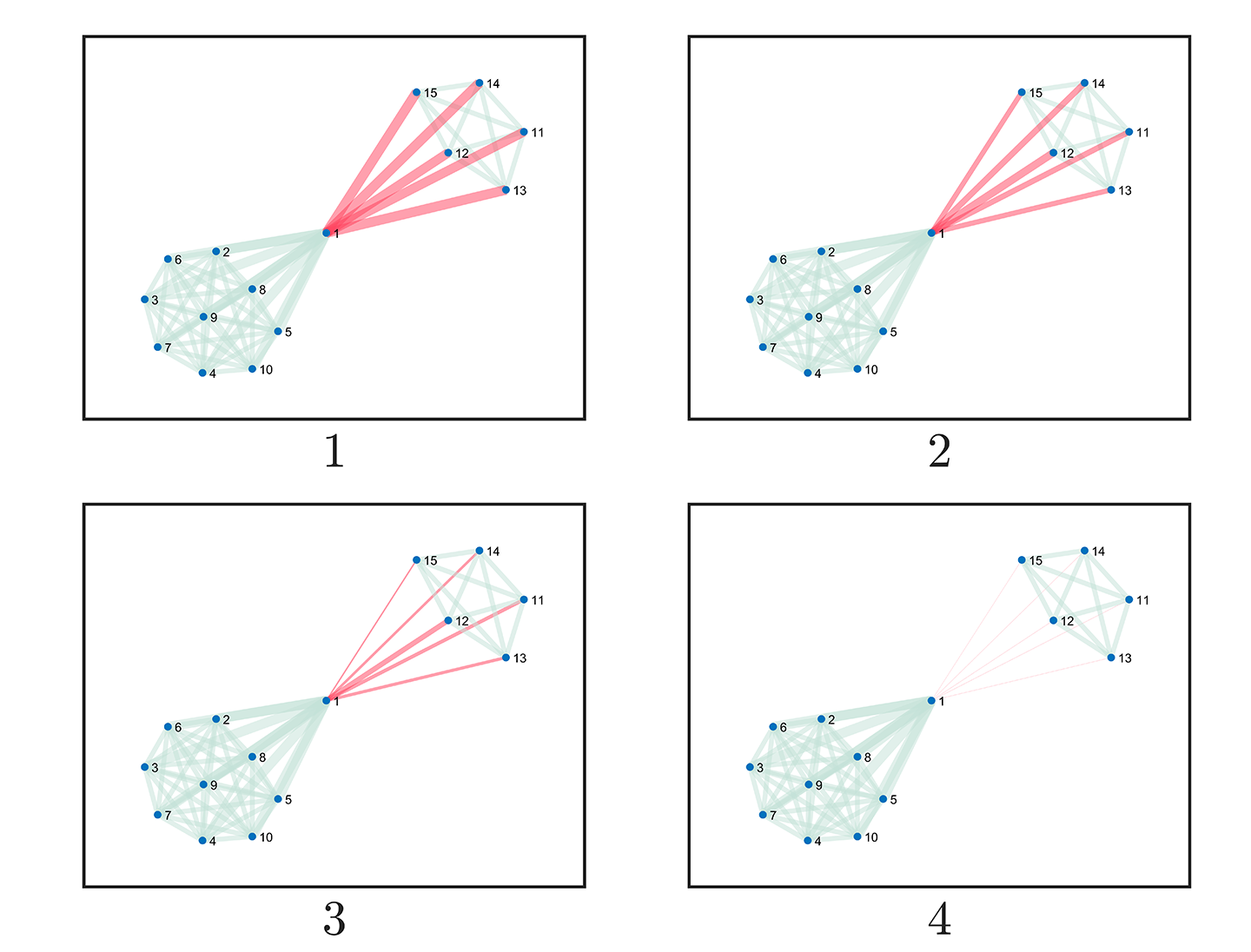}
	\caption{Snapshots of a dynamic company-customer social network model in (\ref{eq:edge_weights}) at four time instants. The edges between utilitarian customers (top right vertices) and the company (center vertex) grow weaker as the social network evolves into segregation with two communities.}
	\label{fig:fig1a}
\end{figure}

\textit{Example 6. Social Media Marketing: }
Consider a social media marketing scenario (e.g. Facebook Business page) where a company is connected with customers. Customers are classified into fans and utilitarian customers \cite{wallace2014likes}. As depicted in Fig. \ref{fig:fig1a}, while fans (bottom left vertices) have a stable connection strength~(i.e.~fixed edge weight) with the company (center vertex), utilitarian customers (top right vertices) have time-varying connection strength~(i.e.~time varying edge weights) with the company due to reasons such as sales events. The variable connection leads to segregation and integration of the company-customer social network. 

\subsection{Related Work}
Echo chambers are studied in various social networks from different modelling perspectives. \cite{cinelli2020echo} studies users' leaning about controversial news by analyzing features of their interaction networks on social media, such as shared links domain and follower relationship. \cite{sasahara2019inevitability} formulates the emergence of echo chambers on Twitter's follower network by introducing a model of social influence and unfriending. Users can change both their opinions and social connections based on the information to which they are exposed. \cite{baumann2020modeling} proposes a group polarization dynamic model from a social psychology perspective. It proposes radicalization as a reinforcing mechanism to drive the evolution of echo chambers with extreme opinions from moderate initial conditions. \cite{ge2020understanding} analyzes the echo chamber phenomenon in e-commerce recommender systems. It illustrates that user interests are self-reinforced through repeated exposure to similar contents and further polarized through the echo chamber formed by such recommendations. \cite{barbera2015tweeting} estimates the ideology political preferences of Twitter users using correspondence analysis and analyzes cross-ideological dissemination of liberals and conservatives. It concludes that the echo chamber effect might be overestimated in social-media usage. \cite{garimella2018political} computes the ideology political leaning of Twitter users based on the news organization they follow. It studies the echo chambers in both spread and consumption of information.

Many other works view echo chambers as a polarized consensus state of a multi-agent system and build upon DeGroot’s well-known model of opinion formation \cite{degroot1974reaching}. \cite{dandekar2013biased} complements a biased assimilation mechanism into DeGroot's model which strengthens the individual's self-opinion and ensures polarization. \cite{liu2020markov} proposes a mechanism considering the co-evolution between information states and network topology. It finds threshold values for the emergence of echo chambers. \cite{de2016learning} models individual users' opinions over time by marked jump diffusion stochastic differential equations, and leverage parameter estimation from historical fine grained event data to identify conditions under which opinions converge to a steady state.


\section{Markov Bridge Model for dynamic social networks}
\label{sec:model}
This section presents a stochastic model to represent the evolution of a social network whose state is fixed at the beginning and at the end. The two fixed states correspond to a segregated social network~(with multiple communities) and a social network which has a single community (i.e.~an integrated network). Thus, the model presented in this section is a useful, intuitive representation of the process of social network segregation. Further, as we show later in Section~\ref{sec:estimation}, the proposed model is easily amenable to Bayesian statistical inference, making it useful in data-driven contexts in computational social science.

\subsection{Time-varying Edge Weight Graph Model}\label{subsec:network_model}
This subsection explains the graph model with time-varying edge weights using a company-customer social network consisting of a company and two types of customers as the graph vertices. The binary classification of customers is motivated from the topological study of Facebook fans in \cite{wallace2014likes}.

The social network at discrete time instant $t$ is modeled by an undirected, weighted graph $G^{(t)}(V, E, w^{(t)})$, with $|V|$ number of agents, $|E|$ number of undirected edges representing their connectivity, and  $w^{(t)}:  E^{(t)} \rightarrow \mathbb{R}^{+}$ representing the weights of the edges (i.e.~the strengths of each connection). 

Let $|V|=n$ be the number of vertices in the network, $M = \{v_1,\cdots, v_m\} \subset V$ be the set of utilitarian customers who are all connected with each other (i.e., form a complete subgraph), $v_{m+1} \in V$ be the company, and $N = \{v_{m+2},\cdots, v_n\} \subset V$ be the set of other customers (fans) that form another complete subgraph. Then, edge weight function $w_{ij}^{(t)}$ between $v_i$ and $v_j$ where $(v_i, v_j) \in E$ is as follows:
\begin{equation} \label{eq:edge_weights}
    w_{ij}^{(t)} = \left\{
        \begin{array}{lr} W_{ij}(t)  & if\; v_i=v_{m+1}\ \&\ v_j \in M \\ 
        &  \vee\; v_j=v_{m+1}\ \&\ v_i \in M \\
        1  & if\; v_i=v_{m+1}\ \&\ v_j \in N \\
        &  \vee\; v_j=v_{m+1}\ \&\ v_i \in N \\
        1  & if\; (v_i, v_j) \in  M\ \&\ v_i \neq v_j \\
        &  \vee\; (v_i, v_j) \in N\ \&\ v_i \neq v_j \\
        \end{array}
    \right.
\end{equation}
where, $W_{ij}(t), \, t = 1, 2, ..$ is the Markov bridge process that we define in Section~\ref{subsec: MB}.

Note that (\ref{eq:edge_weights}) classifies the edge weights into three groups: between company and utilitarian customers, between company and fans, and between two customers of the same type. Note that there is no edge between two customers of different types. The edge weights between company and utilitarian customers is subject to sales events and therefore described as a time-evolving random process $W(t)$~(specified in Section~\ref{subsec: MB}). Other weights are simply set to be one. The simplification is reasonable because fans would be indifferent about sales events and have a more stable relationship with the company. Further, we also assume that the customers of the same type are all connected with each other motivated by the concept of homophily \cite{mcpherson2001birds}. 

\subsection{Markov Bridge Model of Edge Weights} \label{subsec: MB}
 We now propose a Markov bridge model (MB) for the evolution of the weight $W_{ij}(t)$ in the graph. Recall \cite{white2011optimal} a MB is a one-dimensional Markov random field. It is clamped at the beginning and end time point and evolves in between with a three point transition probability $p\{W_{ij}(t) | W_{ij}(t+1), W_{ij}(t-1))\}$. A MB for $W_{ij}(t)$ facilitates modeling a community that separates and then reintegrates with another community in a network. Unlike a Markov chain which enters a state at a geometrically distributed time, a MB enters a state at a fixed deterministic time \cite{5947152}.

We consider $(2T-1)$ time steps as the period between two consecutive sales events. The edge weight $W_{ij}(t)$ between company and utilitarian customers reaches maximum at time $1$ and time $2T-1$ when sales event happens, and decreases to minimum at time $T$ in the middle of two sales events. The process can be described as two consecutive MBs as we explain next. 

The Markov process $W_{ij}(t)$, $t=1,\cdots,2T-1$ takes value in some finite state space $S=\{0, \frac{1}{N_S-1},\ \cdots, \frac{N_S-2}{N_S-1},\,1\}$, which is an arithmetic sequence with $N_S$ elements. The transition matrix of the Markov process is chosen to be a $N_S\times N_S$ row-normalized Toeplitz matrix such that transitions from a given state to neighboring states~(i.e.,~values in $S$ that are closer to the given state) are more likely. Let the entries of the transition matrix be $P_{a, b}=\emph{P}\{W_{ij}(t+1)=b|W_{ij}(t)=a\}$ for all edge weights $W_{ij}$ in (\ref{eq:edge_weights}), where $a, b \in S$ are two states of the social network. 
In this setup, we fix the states at $t=1, T, 2T-1$ of a Markov process - this can be viewed as two sequential MBs: one that starts at time $1$ and another one that starts at time $T$. Both MBs have their starting and end states fixed. The first MB's end state overlaps the second MB's starting state.
The first MB (for each edge)
is initialized as $1$ and
the state at time $T$ is set to be $0$,~i.e. $W_{ij}^1 = 1,\,W_{ij}^T = 0$. Thus, the transition probability of the first MB going from state $a$ to state $b$ is obtained by applying the Bayes rule as follows \cite{white2011optimal}:
\begin{equation}\label{eq_mb_transition}
\begin{split}
B^{c}_{a,b}(t) & = \emph{P}\{W_{ij}(t+1)=b|W_{ij}(t)=a, W_{ij}(T)=0\} \\
 & = \frac{P_{a,b}(P^{T-(t+1)})_{b,c}}{(P^{T-t})_{a,c}}
\end{split}
\end{equation}
for $t=0,\cdots,T-2$, where $c=1$ which is the order of $0$ in the state space.

Likewise, the state of social network is fixed to be $1$ at $2T-1$~(i.e.,~the last time step)
We can then formulate the transition probability of the second MB in a similar manner as follows:
\begin{equation}\label{eq_mb_transition_2}
\begin{split}
B^{c'}_{a,b}(t) & = \frac{P_{a,b}(P^{2T-(t+1)})_{b,c'}}{(P^{2T-t})_{a,c'}}
\end{split}
\end{equation}
for $t=T-1,\cdots,2T-2$, where $c'=N_S$ which is the order of $1$ in the state space. Thus, the dynamics of edge weights is specified by two MBs with transition probability matrices given by (\ref{eq_mb_transition}),  (\ref{eq_mb_transition_2}) and the fixed initial state of the first MB.

\subsection{Graph Clustering Metrics}

The aim of this subsection is to discuss the graph metric called graph conductance that we use to express segregation and set as the state of our model. Graph conductance is a measurement of the level of clustering in a graph and is explained below.

We first define a cut $(S, \overline{S})$ as a partition of the vertices of a graph into two disjoint subsets $S$ and $\overline{S}$. The conductance of a cut $(S, \overline{S})$ in a graph is defined as: 
\begin{equation}\label{eq_conductance_define}
\phi(S) = \frac{\sum_{i\in S}\sum_{j\notin S} w_{ij}}{\min\{a(S), a(V\setminus S)\}} ,\; S\subset V
\end{equation}
where $a(S)=\sum_{i\in S} \sum_{j\in V} w_{ij}$ is the sum of the weights of all edges with at least one endpoint in $S$. Then, given a graph $G$, we define the graph conductance as the minimum conductance over all possible cuts
\begin{equation}\label{eq_conductance_define_2}
\phi(G) = \min_{S\subset V} \phi(S).
\end{equation}

\begin{figure}
	\centering
	\includegraphics[width=0.5\textwidth]{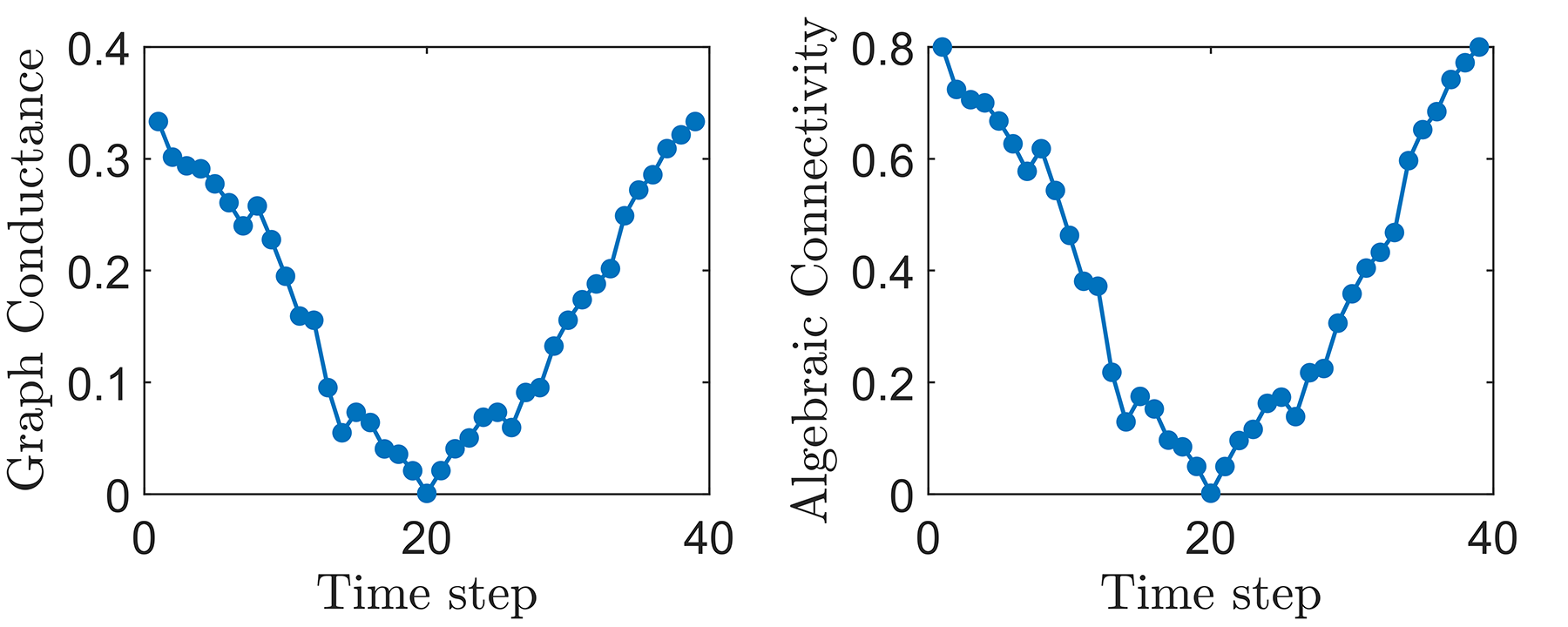}
	\caption{Two metrics (left: graph conductance $\phi(G)$, right: algebraic connectivity $\lambda_2$) that indicate the strength of connectivity between communities in a graph. The figure shows that conductance (the state random variable in our model) resembles other metrics such as the algebraic connectivity.}
	\label{fig:fig2}
\end{figure}

Graph conductance is also related to the algebraic connectivity which is the second-smallest eigenvalue of the Laplacian matrix of $G$. Algebraic connectivity is used in many results in spectral graph theory such as Cheeger's inequality~\cite{chung1997spectral}. 
The derivation of algebraic connectivity can be found in \cite{de2007old}.
The weighted adjacency matrix $A^{(t)}$ of the graph is given by
\begin{equation}
    A^{(t)}_{ij} = w_{ij}^{(t)},\
\end{equation}
The degree matrix $D^{(t)}$ is given by
\begin{equation}
    D^{(t)}_{ij} = \left\{
        \begin{array}{lll}
        \sum_{k}^{} A^{(t)}_{ik} & & \text{if}\; i=j \\
        0 & & \text{otherwise}
        \end{array}
    \right.
\end{equation}
The Laplacian $L^{(t)}$ of the graph is given by
\begin{equation}
    L^{(t)} = D^{(t)} - A^{(t)}
\end{equation}
Fig. \ref{fig:fig2} shows the variation of both graph conductance and algebraic connectivity follow a similar dynamics. This implies that an estimate of the graph conductance also serves as a proxy for the algebraic connectivity under our model.

\section{Bayesian Estimation of Graph Metrics}
\label{sec:estimation}

Section~\ref{sec:model} presented a MB for a social network segregation. A natural question is: \textit{assuming the MB with known parameters, how can one estimate the level of segregation in a data driven manner?} An answer to this question is useful in computational social science and network science that deal with large scale, partially observable (via noisy samples) social networks. As a solution, we propose a Bayesian filtering method based on the proposed segregation model. 

\subsection{Measuring Conductance via Sampled Edges} \label{subsec:measurement_noise}
This subsection discusses our criteria for obtaining a noisy estimate of the conductance of the underlying dynamic graph (explained in Section~\ref{sec:model}) using a sampled subgraph. We demonstrate that the sampling noise can be approximated as a Gaussian noise from the central limit theorem.

We assume that $\gamma N$ of the total $N$ edges are uniformly sampled and observed at each time $t$~(random sampling of edges has been used widely in literature in statistical estimation tasks e.g.~\cite{nettasinghe2019your, nettasinghe2019friendship, nettasinghe2019maximum}). $\gamma $ is a fixed ratio in $(0, 1]$. The observed graph conductance $\phi(\tilde{G}^{(t)})$ is computed from the partially sampled graph $\tilde{G}^{(t)}$ at time $t$. Graph conductance is a static function of edge weights,
\begin{equation}\label{eq_static_function}
    \phi(G^{(t)}) = f(w^{(t)}_1,\cdots,w^{(t)}_N)
\end{equation}
And the observed graph conductance is the same function of sampled edge weights,
\begin{equation}
    \phi(\tilde{G}^{(t)}) = f(w^{(t)}_{i_1},\cdots,w^{(t)}_{i_{\gamma N}})
\end{equation}
where $i_1,\cdots,i_{\gamma N}$ are sampled from $1,\cdots,N$ with equal probabilities. 
From (\ref{eq_static_function}), it follows straight forwardly that graph conductance as a static function of the edge weights, follows the same MB dynamics. For the rest of the paper, we denote $\phi(G^{(t)})$ and $\phi(\tilde{G}^{(t)})$ as $\phi^{(t)}$ and $\tilde{\phi}^{(t)}$ respectively.

\begin{figure}
	\centering
	\includegraphics[width=0.5 \textwidth]{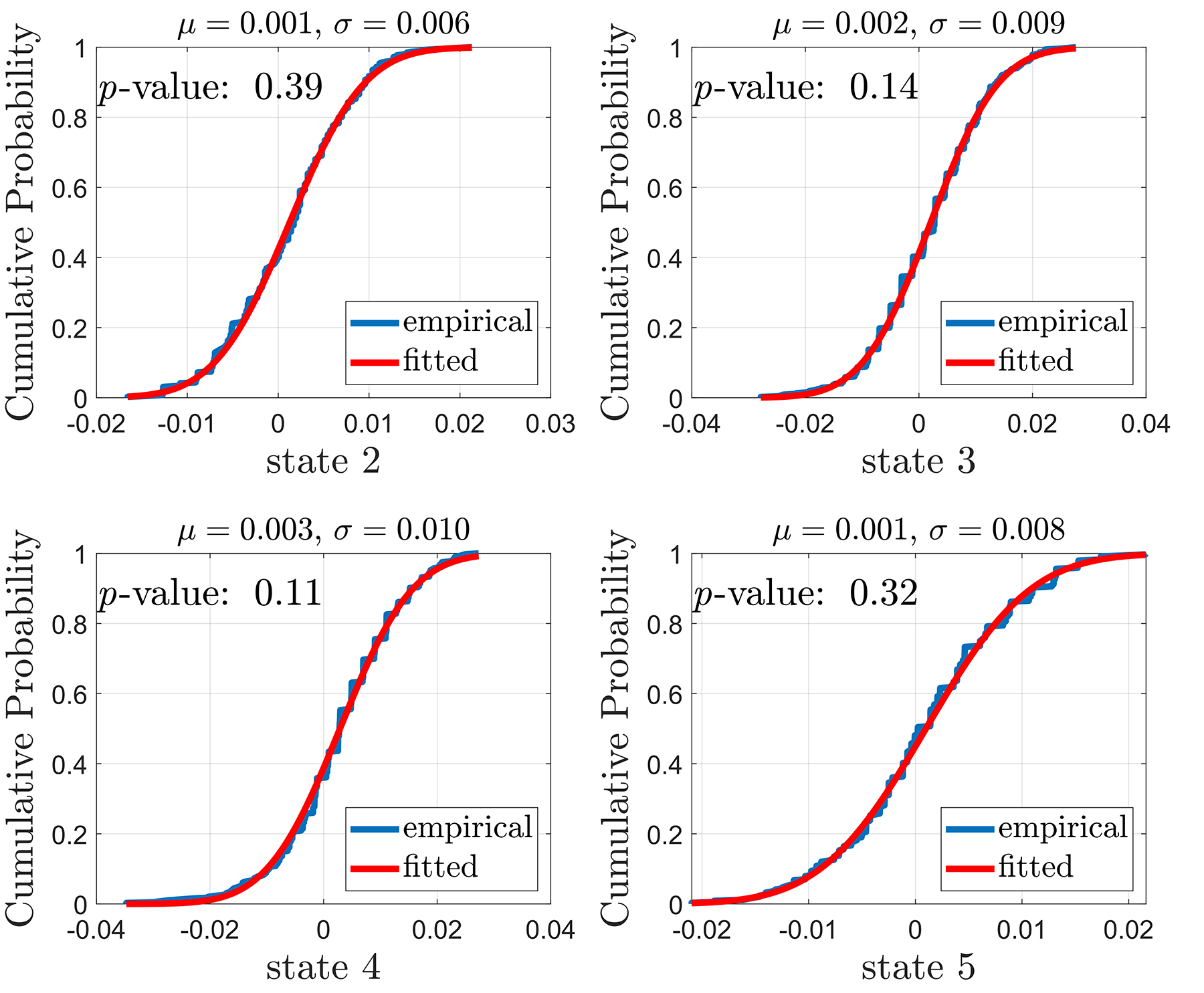}
	\caption{This figure shows that the empirical CDF of the sampling noise for graph conductance can be fitted as a Gaussian distribution. It contains the empirical and fitted cumulative probability at four states in a simulation of 6-state model. The \emph{p}-value of KS test is above 0.05 significance level and therefore the Gaussian distribution null hypothesis is unrejectable.}
	\label{fig:fig1b}
\end{figure}

To estimate the observation probabilities $p(\tilde{\phi}^{(t)}|\phi^{(t)}=j)$, we use a Monte Carlo simulation to obtain sample trajectories of the $(2T-1)$ step graph evolution and compute the empirical CDF of noise of the conductance computed from the partial observation~i.e.~the CDF of the difference between the estimated conductance $\gamma\tilde{\phi}^{(t)}$ and the true conductance $\phi^{(t)}$. Fig. \ref{fig:fig1b} shows that the observation noise is approximately~(in the sense of Kolmogorov-Smirnov test) a Gaussian distribution~i.e. 
\begin{equation}\label{eq_ecdf}
p(\tilde{\phi}^{(t)}|\phi^{(t)}=i) \sim \mathcal{N}(\gamma\tilde{\phi}^{(t)} - \phi^{(t)}|\mu^{(t)}, {(\sigma^{(t)})}^2)
\end{equation}

The normal distribution form of the observation noise~(\ref{eq_ecdf}) can also be viewed as a consequence of the central limit theorem: since we are sampling i.i.d edge sequences from the social network and approximate the graph conductance using the average of their weights, it follows from the central limit theorem that the sample mean~(scaled by the square root of the number of samples) converges in distribution to a Gaussian distribution centered around the true state.

\subsection{Hidden Markov Bridge Filter} \label{subsec:filter}
In this subsection, we aim to estimate the segregation level of a social network by computing the posterior probability of the graph conductance given its sampled observation (the conductance computed from sampled edges of the graph). Section~\ref{subsec:measurement_noise} exploits the Gaussian approximation of measurement noise and we propose a Hidden Markov bridge (HMB) filter here for recursively tracking the state of the graph conductance. HMB filter is a generalization of the time-homogeneous HMM filter~\cite{krishnamurthy2016partially} and have been widely used in signal processing methods for target tracking~\cite{fanaswala2011destination, fanaswala2012detection, fanaswala2015spatiotemporal}).  

Suppose that the MB process ${\Phi} = \{\phi^{(1)}, \cdots, \phi^{(t)} \}$ is observed via the observation process $\tilde{\Phi} = \{\tilde{\phi}^{(1)}, \cdots, \tilde{\phi}^{(t)} \}$. Assume that the observation at time $t$ given the state $\phi^{(t)}$ is conditionally independent of $\phi^{(\tau)}$ and $\tilde{\phi}^{(\tau)}$, $\tau \neq t$. This conditional independence implies that
\begin{equation}\label{eq_hmb_condition}
\emph{P}(\tilde{\phi}^{(1)},\cdots,\tilde{\phi}^{(t)}|\phi^{(1)},\cdots,\phi^{(t)}) = \prod_{k=0}^{t}\emph{P}(\tilde{\phi}^{(k)}|\phi^{(k)})
\end{equation}

The process $\tilde{\Phi}$ is called a Hidden Markov bridge (HMB) because the property (\ref{eq_hmb_condition}) is analogous to the assumption made for Hidden Markov Model (HMM). Consider the HMB $\tilde{\Phi}$ with state $\Phi$, known MB transition probability (\ref{eq_mb_transition}), and pre-computed observation probability (\ref{eq_ecdf}). The posterior probability can be evaluated recursively via Bayes' rule
\begin{equation}\label{eq_posterior}
q_j(t+1) = \frac{p(\tilde{\phi}^{t+1}|\phi^{t+1}=j)\sum_{i=1}^{\Phi} B_{i,j}^k(t)q_i(t)}{\sum_{l=1}^{\Phi} p(\tilde{\phi}^{t+1}|\phi^{t+1}=l)\sum_{i=1}^{\Phi} B_{i,j}^k(t)q_i(t)}
\end{equation}
as shown in \cite{stamatescu2017track}.

\section{Numerical Examples on Social Media Marketing and Twitter Political Retweets}
In this section, we numerically illustrate that the proposed HMB filter (Section~\ref{sec:estimation}) outperforms (in terms of mean-squared error) the widely used HMM filter for estimating the level of segregation on synthetic data. This highlights how the proposed model and filtering method can be useful in estimating the level of segregation with a better accuracy compared to the baseline method of HMM filtering. We also evaluate the proposed model on a public Twitter election dataset. 

\subsection{Simulation on Weighted Customer-Merchant Graph}
In this subsection, we consider a social media marketing scenario where the connection between company and customers is modeled as a MB. A HMB filter is implemented to estimate the inter-community distance based on sampled observation of single edge weights and an additive Gaussian noise. It outperforms a hidden Markov chain filter regarding the mean-squared error.

\subsubsection{Simulation setup} 
We consider a company-customer network of 10 utilitarian customers, 20 fans, and 1 company as discussed in Section~\ref{sec:model} for $2T-1$ time steps ($T=20$). The state space of the weight of each edge between utilitarian customers and company is an arithmetic sequence $\big[1, \frac{N_S-2}{N_S-1}, \cdots, \frac{1}{N_S-1}, 0\big]$. The weight evolves according to a transition matrix which is a Toeplitz matrix. Each descending diagonal from left to right is constant: $\big[(\frac{1}{4})^{\scriptstyle{N_S-1}},\cdots,1,\ \cdots,(\frac{1}{4})^{\scriptstyle{N_S-1}} \big]$. Each row vector of the Toeplitz matrix is normalized so that the row elements add up to $1$. We then implement the HMB filter in assuming the measurement noise is Gaussian with the empirically estimated mean and covariance in Section~\ref{subsec:measurement_noise}. To assess the performance, we compare the mean-squared error of the HMB filter with a HMM filter that assumes the underlying process is a Markov chain (instead of a MB). 

\label{sec:simulation}
\begin{figure}
	\centering
	\includegraphics[width=0.5\textwidth]{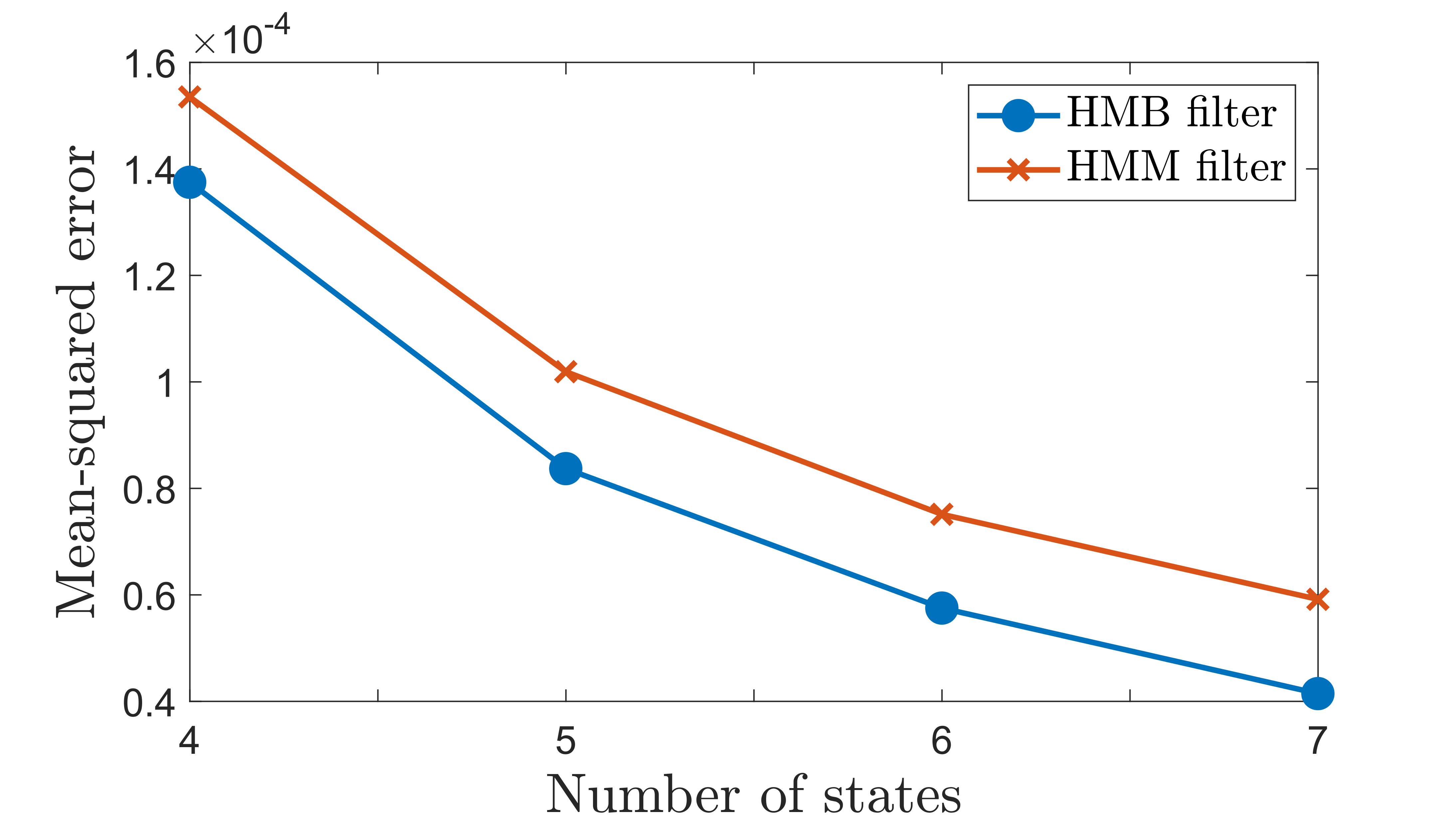}
	\caption{Mean-squared error of the proposed Hidden Markov bridge (HMB) filter compared with a Hidden Markov Model (HMM) filter in the company-customer marketing network simulation. HMB filter outperforms HMM filter by approximately $20\%$, which indicates its better prediction ability of segregation in social networks.}
	\label{fig:fig3}
\end{figure}

\subsubsection{Numerical Results of Filters} 
Fig.~\ref{fig:fig3} depicts the results obtained using the above simulation setup. Results show that the proposed HMB filter outperforms the HMM filter for all considered number of states ($N_S$ values). Thus, the numerical results indicate that the proposed Bayesian filter is capable of accurately estimating the level of segregation in a company-consumer network from noisy sampled edges.

\subsection{Hidden Markov Bridge Model of Political Polarization during 2020 Election}
In this subsection, we propose that the MB model is sociologically beneficial for predicting the emergence and segregation level of echo chambers on a social network. We justify our conclusion on a real-world Twitter dataset where a polarization score is defined on the Twitter retweet network. We determine the model parameters (i.e. the transition matrix) from maximum likelihood algorithm which is derived in Appendix \ref{sec:EM}. We apply the HMB filter to estimate the polarization score. The filter's estimation accuracy outperforms a HMM filter regarding mean-squared error. We also verify that the observation noise can be approximated as a Gaussian distribution based on statistical hypothesis testing. 

\subsubsection{Construction of Retweet Network and Polarization Score}
\label{subsubsec:retweet network}

\begin{table}
\centering
\begin{tabular}{ C{0.2\textwidth}C{0.11\textwidth} }
\textbf{account name}& \textbf{political party} \\ 
\hline
@realDonaldTrump & R \\
@GovBillWeld & R \\
@MarkSanford & R \\
@WalshFreedom & R \\
@JohnDelaney & R \\
@AmbassadorRice & R \\
@TrumpWarRoom & R \\
@TeamTrump & R \\
@JoeBiden & D \\
@CoryBooker & D \\
@GovernorBullock & D \\
@SenKamalaHarris & D \\
@BernieSanders & D \\
@SenWarren & D \\
@marwilliamson & D \\
@AndrewYang & D \\
\end{tabular}
\caption{A sample of Election-related Twitter accounts tracked in the dataset.} \label{table: accounts}
\end{table}

\begin{figure}
	\centering
	\includegraphics[width=0.5 \textwidth]{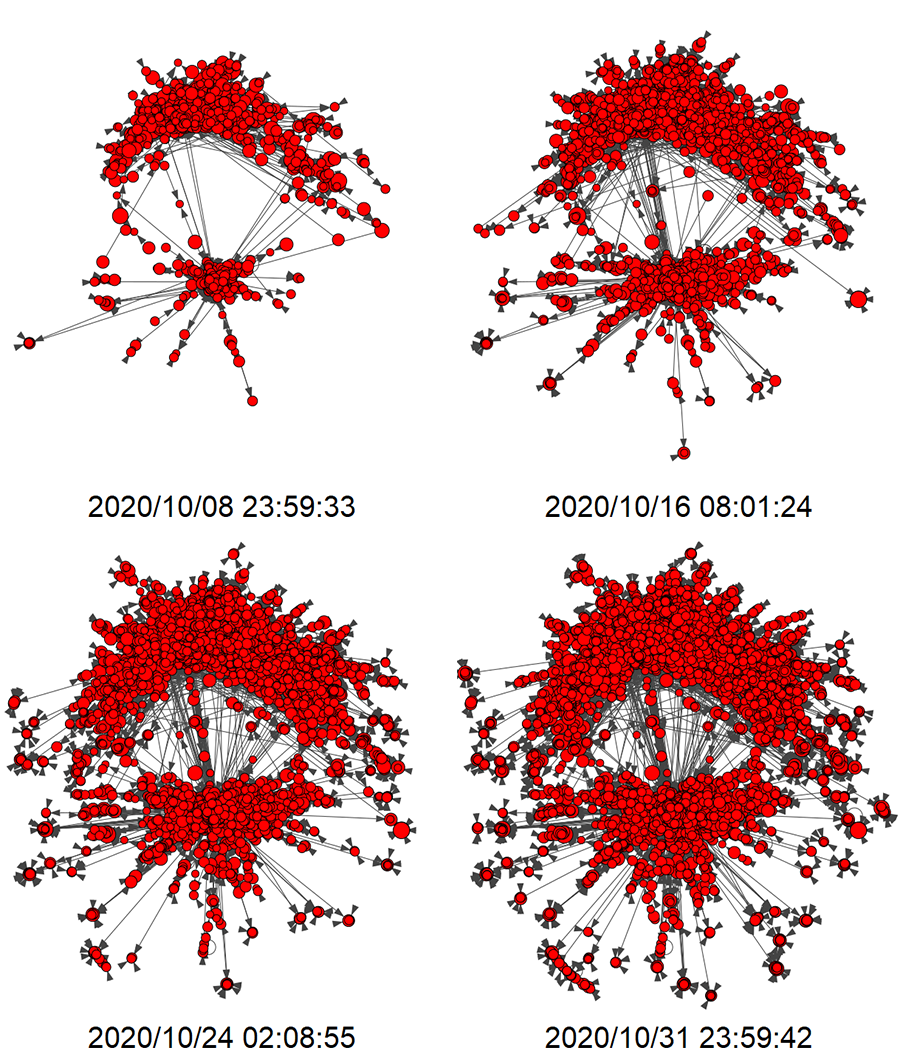}
	\caption{Snapshots of a dynamic retweet network with two communities at 4 different timestamps before the Nov. 3 presidential election. The ratio of intra-community connections and total connections (sum of intra- and inter-community connections) evolves and is modeled as a Markov bridge in (\ref{eq_polarization_score}). The graph is laid out using DrL (same as Fig.\ref{fig:fig1c}).}
	\label{fig:fig4}
\end{figure}

We leverage a publicly available dataset which encompasses 240 million tweets related to the 2020 U.S. presidential election. The dataset captured tweets with specific user mentions and accounts (57 in total) that are tied to president candidates and politicians. A sample of such accounts is shown in Table \ref{table: accounts}. The column of political party denotes the party which this account belongs to (D-democratic, R-republican). We select and sample tweets from Oct. 1 to Nov. 1, spanning a 30-day period before the election day (Nov. 3) which is about 7 million tweets in total. See Appendix \ref{sec:Twitter data} for data collection and sampling procedure. 

From this subset of tweets, we constructed a dynamic retweet graph $G^{(t)}(V, E^{(t)}),\; t=1,\cdots,30$, where the nodes represents $|V|$ Twitter accounts, directed and unweighted edges $E_{ij}^{(t)}$ from node $i$ to node $j$ if user $j$ retweets a message\footnote{Here we do not consider "quote tweets" (retweet with a comment added) to avoid the use of "quote tweets" for ironic or criticizing purposes.} originally posted by user $i$ on day $t$. We filtered out nodes with out-degree fewer than 2 (which means they only retweets once during the 30-day period). In summary, the retweet network has $|V|=1399644$ vertices and $\sum_{t=1}^{30} |E^{(t)}| = 5047498$ edges during the specified period. In Fig. \ref{fig:fig4}, we take a sample of the retweet graph and plot 4 snapshots of its largest weakly connected component. It clearly shows the pattern of partitioning into two polarized community. 

To study the evolution of political opinion polarization during the 30-day period, we define a temporal variable named polarization score. We select users who have retweeted election-related accounts from both political parties. We estimate these users' political leaning (interchangeably referred to as ideology) as follows: Every retweet to accounts from either political party increases the count for that side by +1. The user's political leaning is classified as the side with more accumulated retweets. We dismiss the users whose retweets are equally sourced from two political parties. After this ideology classification, we denote $D = \{v_1,\cdots, v_m\} \subset V$ be the set of $m$ left-leaning users, and $R = \{v_{m+1},\cdots, v_{m+n}\} \subset V$ be the set of $n$ right-leaning users. We then define the polarization score of the retweet network on day $t$ as the ratio of the amount of intra-ideological retweets (e.g. user from $D$ retweets an election-related account who is from the Democratic party) to the amount of intra-ideological retweets plus the amount of cross-ideological retweets (e.g. user from $D$ retweets an election-related account who is from the Republican party) on day $t$. We collected 28 such 30-day sequences of polarization scores, which are used for training the transition matrix and computing the empirical distribution of the observation error.
\begin{equation}
\label{eq_polarization_score}
    y^{(t)} = \frac{|E_{ij}^{(t)}|_{(v_i,v_j) \in D \; \vee\; (v_i,v_j) \in R}}{|E_{ij}^{(t)}|_{}\forall (v_i, v_j)}
\end{equation}


\subsubsection{Hidden Markov Bridge Filter Estimation Results}
We formulate $x^{(t)}$ as a hidden Markov bridge with $y^{(t)}$ as its observation. The hidden states form an arithmetic sequence with maximum and minimum corresponding to those of the observation. We anticipate a high-level opinion polarization near or on election day since people are required to vote to one candidate from one party. Therefore we set the destination, i.e. the final state to be the maximal state. 

\begin{figure}
	\centering
	\includegraphics[width=0.5 \textwidth]{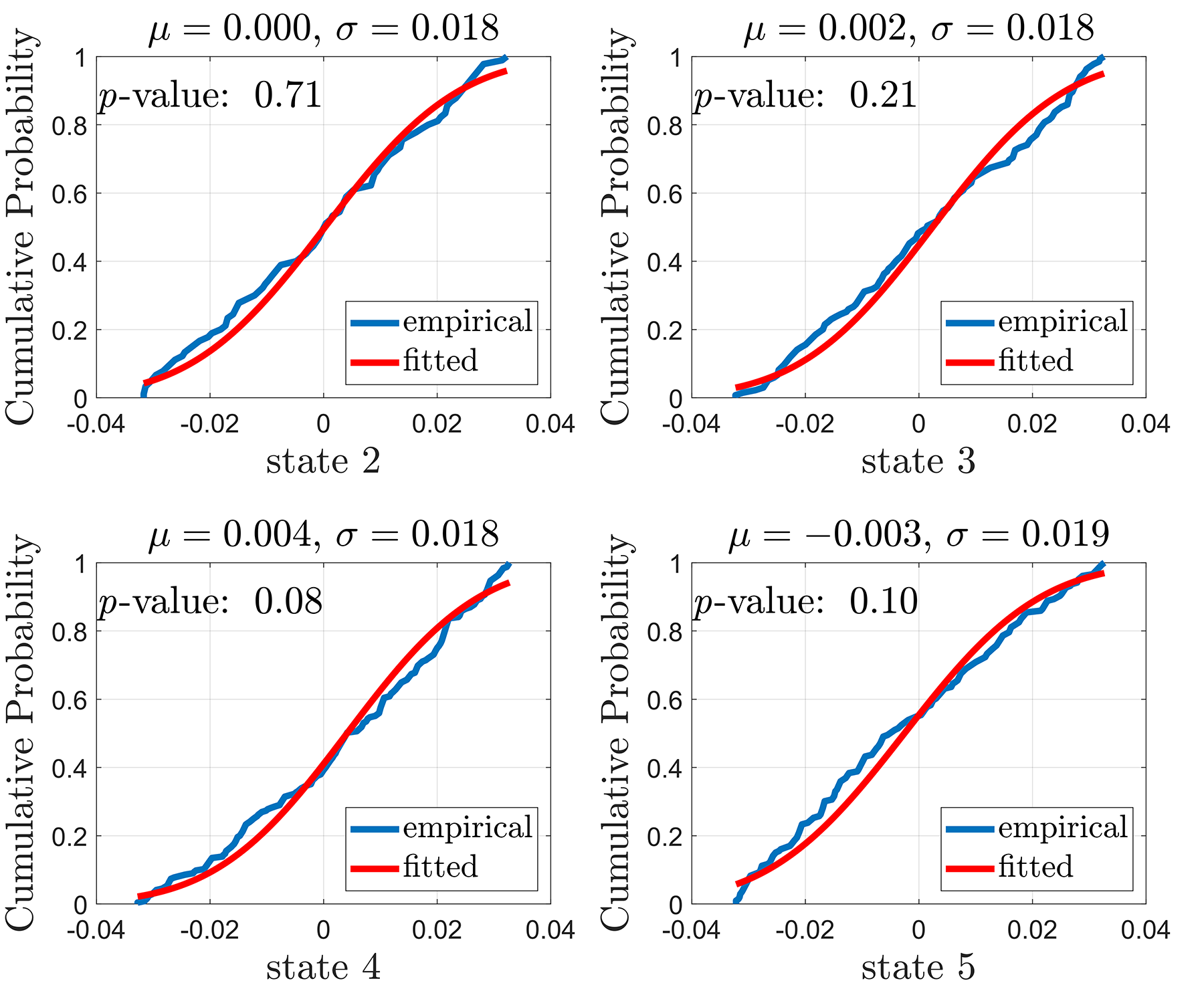}
	\caption{The empirical CDF of the observation noise for polarization score on the Twitter dataset and the CDF of a Gaussian distribution of four hidden states (state 2, 3, 4, 5) in the 6-state model. The \emph{p}-value of KS test indicates that the observation noise can be approximated by Gaussian noise.}
	\label{fig:fig8}
\end{figure}

$x^{(t)}$ follows transition probabilities as depicted by (\ref{eq_mb_transition}) with $c$ as the maximal state, and transition matrix $P$ derived from maximum likelihood algorithm in Appendix \ref{sec:EM}. The observation noise is hypothetically verified to follow a Gaussian distribution as shown in Fig. \ref{fig:fig8}. 

\begin{figure}
	\centering
	\includegraphics[width=0.5\textwidth]{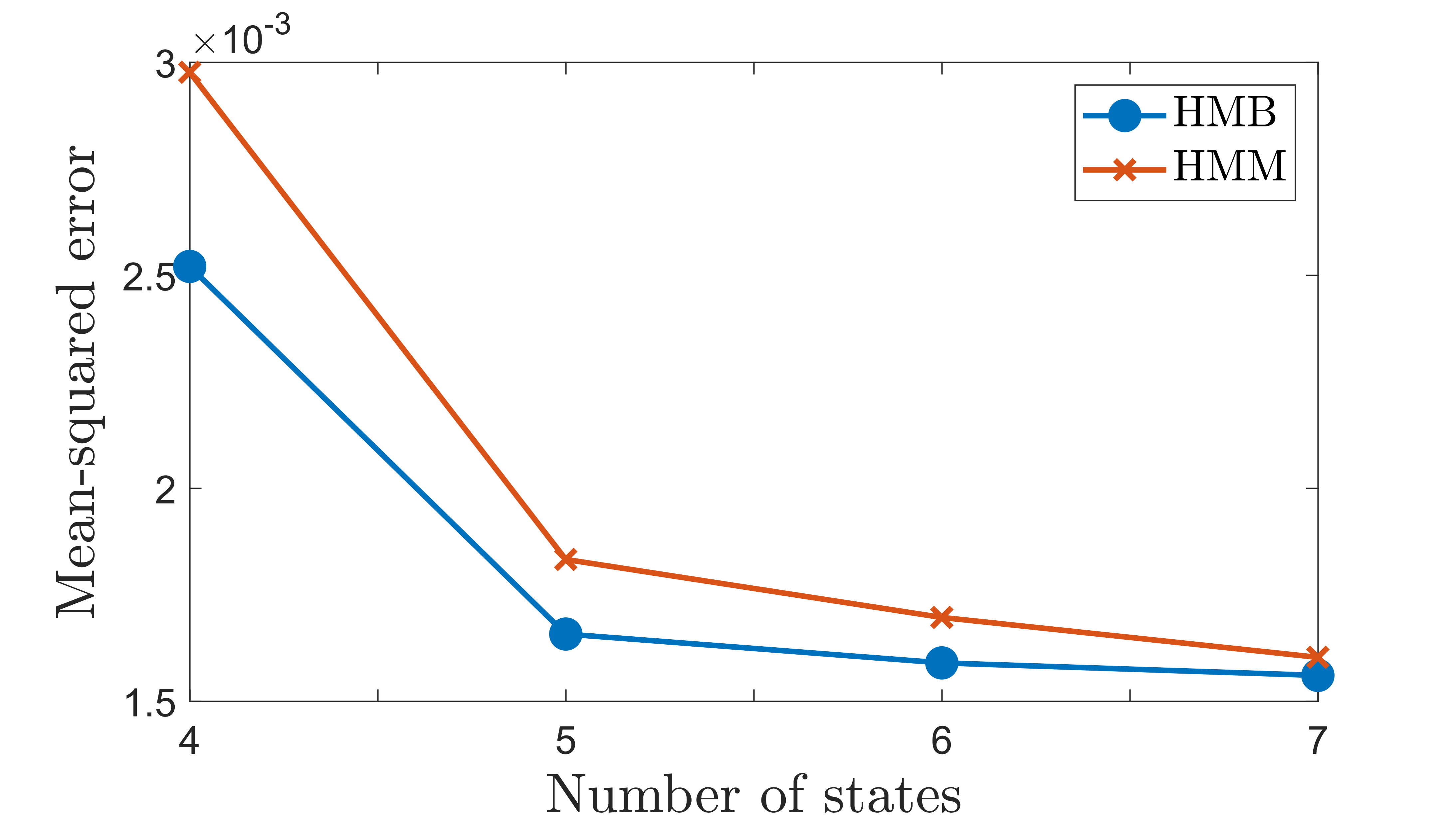}
	\caption{Mean-squared estimation error of the polarization score on Twitter dataset. It shows that the proposed Hidden Markov bridge (HMB) filter outperforms a Hidden Markov Model (HMM) filter in estimating the level of polarization in a social network.}
	\label{fig:fig9}
\end{figure}

To assess the performance of the proposed filter in (\ref{eq_posterior}), we compare its mean-squared error with a HMM filter. Results show that the proposed HMB filter outperforms the HMM filter for all considered number of hidden states and reduces the mean-squared error by 10\% (Fig. \ref{fig:fig9}). 


\subsection{Discussion on the Experiment with Real-world Twitter Data}
The numerical evaluation section explored modeling and estimation of segregation in social networks. For the first social media marketing example, we proposed a tractable model for segregation based on Markov bridge processes of the connection strength between company and customers. We also proposed a Bayesian filter (named HMB filter) to estimate the level of segregation (as measured by graph conductance) by polling some random pairs of neighbors (i.e.~edges) in the network. 

To validate the proposed model with real world data, our second opinion polarization example studied Twitter users' retweet behavior before 2020 presidential election. We anticipated a high level of polarization near the election and modeled the portion of users' retweets to someone of same political ideology as a Markov bridge process. Extension of this study could help answer the following two research questions:

(1) As has been demonstrated in \cite{barbera2015tweeting}, are liberals more likely than conservatives to involve in cross-ideological dissemination? 

(2) In addition to political events such as election, would other topics elicit similar polarized echo chambers in online conversation?

Nevertheless, one major limitation of the study is that we equally weight each retweet. As further improvement, we could consider the following four aspects: 

(1) Consider tweet's temporal effect. We could formulate the influence of tweets to decrease exponentially after posted so that newer tweets would be more informative and representative of user's political opinion. While many tweets spread within a limited time window, some could achieve high virality. Therefore we could assign different decay rate for each tweet as well.

(2) Define user's political ideology as a time-evolving variable instead of a constant. Although a user is typically unchanged as to his favorable party, his political leaning might change due to personal experience and external information. We could adapt the correspondence analysis or the maximum likelihood logit model in \cite{barbera2015birds} to obtain user' political ideology at different times and assign political scores accordingly to the user-related tweets.

(3) Explore the content and sentiment of tweets. It is found in \cite{tsugawa2015negative} that tweets with negative emotion tend to get reposted more rapidly and frequently than positive and neutral messages. Take one step forward, we could decompose each tweet's popularity into its political effect and content effect, and mitigate the unbalance of different sentiments or wording among different tweets.

(4) Combine information from multi-level social graphs. In addition to retweet network, we could construct other types of social network from Twitter data, including mention, follow and comment. We could certainly analyze the trend of polarization separately on these networks and average the results, however, it is more promising to construct a multi-level heterogeneous graph to learn a structural representation of polarization.

\section{Conclusion}
This paper studied the sociological phenomena of segregation and echo chambers in social networks. We proposed a Markov bridge dynamics based model for evaluating the interaction between customers and company in a social media marketing scenario. We then justified the model by looking at the evolution of political opinion polarization on a real-world Twitter dataset. We formulated an additive Gaussian measurement noise model for the Markov bridge, derived the EM algorithm for estimating parameters of the hidden Markov bridge model, and proposed a Hidden Markov Bridge filter to estimate the state of segregation and echo chambers based on sample of the social network. The numerical results indicated that our filter outperforms time homogeneous filters such as a HMM filter. 

Future directions of this work include further improving the accuracy of the proposed method using different sampling methods based on friendship paradox~(e.g.~\cite{nettasinghe2019your, nettasinghe2019friendship, nettasinghe2019maximum}); enriching this framework to handle more sophisticated network topologies such as heterogeneous graphs; incorporating the hidden Markov bridge model with generation models to forecast opinion dynamics (e.g. \cite{de2016learning, amati2016twitter, ten2014modelling}).


%



\appendices
\section{Twitter Dataset Details}
\label{sec:Twitter data}
The Twitter dataset was provided by \cite{chen2020election2020} and publicly available\footnote{The dataset website: https://github.com/echen102/us-pres-elections-2020}. The authors used Twitter’s streaming API through Tweepy and kept track of tweets with specific keyword mentions and accounts related to the 2020 US presidential election since May 2019. The data contains approximately $1\%$ stream of all tweets in real-time.

In obedience to Twitter’s Developer Agreement \& Policy, only tweet’s Tweet ID is shared. The Tweet ID is preserved in text files in temporal order. We used Doc-Now’s Hydrator\footnote{https://github.com/DocNow/hydrator} to retrieve the tweet objects\footnote{https://developer.twitter.com/en/docs/twitter-api/v1/data-dictionary/object-model/tweet} with full tweet payloads including the tweet poster, content, timestamp, and the author who is retweeted from.

To reduce the amount of data, we use systematic sampling, which means we pick every $n^{th}$ tweets ($n=20$) to comprise the data used in the research. For each tweet, we keep its poster, author and timestamp if it is a retweet (if the tweet object has a 'retweeted\_status' attribute). 

\section{EM Algorithm for Hidden Markov bridge parameters}
\label{sec:EM}
For real-life application of the Markov bridge model proposed in section \ref{sec:model}, we need to obtain the model parameters which are useful for filtering and forecasting the segregation state of the social network. This section presents the EM algorithm which serves the purpose of finding the maximum-likelihood estimate of the parameters. 

\subsection{Forward-Backward Smoothing Algorithm for Hidden Markov Bridge}
In this subsection we derive the forward-backward (also named Baum-Welch) algorithm for smoothing the hidden Markov bridge (HMB) model. Consider HMB model with parameter $\theta=(S, P, O)$ where $S$ is a $X$-state Markov chain with transition matrix $P=(P_{a,b}),\, a,b\in S=\{s_1,\cdots,s_X\}$, and $O$ is the HMB emission probability function. The HMB model has unknown state sequence $X^{(T)}=(x^{(1)}, \cdots, x^{(T)})$ and observation sequence $Y^{(T)}=(y^{(1)}, \cdots, y^{(T)})$. We know the destination of the state sequence $X^{(T)}$ is $c\in S$. 

We first go through the forward procedure by defining
\begin{equation}
    \alpha_{\theta}^{(t)}(a) = P(x^{(t)}=a, y^{(1)}, \cdots, y^{(t)}|\theta) = P(x^{(t)}=a, Y^{(t)}|\theta)
\end{equation}
which is the probability of seeing the partial sequence $(y^{(1)}, \cdots, y^{(t)})$ and ending up in state $a$ at time $t$.

The backward procedure is similar by defining a backward variable
\begin{equation}
    \beta_{\theta}^{(t|T)}(a) = P(y^{(t+1)},\cdots,y^{(T)}|x^{(t)}=a)
\end{equation}

We now define
\begin{equation}
    \gamma_{\theta}^{(t)}(a) = P(x^{(t)}=a, Y^{(T)}|\theta) 
\end{equation}
which is the probability of being in state $a$ at time $t$ for the observation sequence $(y^{(1)}, \cdots, y^{(T)})$. It can be derived in terms of $\alpha_{\theta}^{(t)}$ and $\beta_{\theta}^{(t|T)}$
\begin{equation}
    \gamma_{\theta}^{(t)}(a) = \frac{\alpha_{\theta}^{(t)}(a) \beta_{\theta}^{(t|T)}(a)}{\sum_{b=s_1}^{s_X} \alpha_{\theta}^{(t)}(b) \beta_{\theta}^{(t|T)}(b)}
\end{equation}
We also define
\begin{equation}
    \gamma_{\theta}^{(t)}(a, b) = P(x^{(t)}=a, x^{(t+1)}=b, Y^{(T)}|\theta)
\end{equation}
which is the probability of being in state $a$ at time $t$ and being in state $b$ at time $t+1$. This can be expressed in terms of $\alpha_{\theta}^{(t)}$ and $\beta_{\theta}^{(t|T)}$
\begin{equation}
\begin{split}
    & \gamma_{\theta}^{(t)}(a, b) =  \\
    & \frac{\alpha_{\theta}^{(t)}(a) B_{a,b}(t) O(y^{(t+1)}|x^{(t+1)}=b) \beta_{\theta}^{(t+1|T)}(b)}{\sum_{a=s_1}^{s_X} \sum_{b=s_1}^{s_X} \alpha_{\theta}^{(t)}(a) B_{a,b}(t) O(y^{(t+1)}|x^{(t+1)}=b) \beta_{\theta}^{(t+1|T)}(b)}
\end{split}
\end{equation}

\subsection{Maximum Likelihood Estimation Algorithm}
We assume the HMB model is observed in Gaussian noise, i.e.~the emission probability follows a zero-mean Gaussian distribution.
\begin{equation}
    y^{(t)} = x^{(t)} + v^{(t)}, \; v^{(t)}\sim N\big(0, \sigma^2\big) 
\end{equation}

The E-step of the EM algorithm finds the expected value of the complete-data log-likelihood with respect to the unknown state $X^{(T)} = (x^{(1)}, \cdots, x^{(T)})$ given the observation $Y^{(T)} = (y^{(1)}, \cdots, y^{(T)})$ and the current parameter estimates $\theta^{I} = (S, P, \sigma)$. This log-likelihood is defined as the Q function
\begin{equation}
    Q(\theta, \theta^{I}) = E\{\log {P(Y^{(T)}, X^{(T)}|\theta)}|Y^{(T)}, \theta^{I}\}
\end{equation}
The second step (the M-step) of the EM algorithm is to maximize the expectation we computed in the first step. That is, we find
\begin{equation}
    \theta^{I+1} = \argmax_{\theta} Q(\theta, \theta^{I})
\end{equation}

The joint probability of states $X^{(T)}$ and observations $Y^{(T)}$ given the parameter is formulated as
\begin{equation}
\begin{split}
    &\log {P(Y^{(T)}, X^{(T)}|\theta)} \\
    &= \log {\prod_{t=1}^{T} P(y^{(t)}|x^{(t)})}P(x^{(t)}|x^{(t-1)},x^{(T)}=c)\\
    &= \sum_{t=1}^{T} \log {P(y^{(t)}|x^{(t)})} + \log{P(x^{(t)}|x^{(t-1)}, x^{(T)}=c)} \\
    &= \sum_{t=1}^{T}\sum_{a=s_1}^{s_X} I(x^{(t)}=a) \log{P(y^{(t)}|x^{(t)}=a)} +\\
    & \sum_{t=1}^{T}\sum_{a=s_1}^{s_X}\sum_{b=s_1}^{s_X} \big[ I(x^{(t)}=a,x^{(t+1)}=b) \\ 
    & \log{P(x^{(t+1)}=b|x^{(t)}=a, x^{(T)}=c)} \big]\\
    &= \sum_{t=1}^{T}\sum_{a=s_1}^{s_X} I(x^{(t)}=a) \big[\log{(\frac{1}{\sqrt{2\pi}\sigma})} - \frac{{(y^{(t)}-a)}^2}{2\sigma^2} \big] +\\
    & \sum_{t=1}^{T}\sum_{a=s_1}^{s_X}\sum_{b=s_1}^{s_X} I(x^{(t)}=a,x^{(t+1)}=b) \log{B_{a,b}^{c}(t)} \\
\end{split}
\end{equation}

The Q function can then be simplified as
\begin{equation}
\begin{split}
    Q(\theta, \theta^{I}) = -\frac{T}{2} \log{\sigma^2} - \frac{1}{2\sigma^2} \sum_{t=1}^{T}\sum_{a=s_1}^{s_X}{(y^{(t)}-a)}^2 \gamma^{(t)}_{\theta^{I}}(a) \\
    + \sum_{t=1}^{T}\sum_{a=s_1}^{s_X}\sum_{b=s_1}^{s_X} \gamma^{(t)}_{\theta^{I}}(a,b) \log{B_{a,b}^{c}(t)} \\
    = \cdots + \sum_{t=1}^{T}\sum_{a=s_1}^{s_X}\sum_{b=s_1}^{s_X} \gamma^{(t)}_{\theta^{I}}(a,b) \log{P_{a,b}\frac{{(P^{T-(t+1)})}_{b,c}}{{(P^{T-t})}_{a,c}}}
\end{split}
\end{equation}
Recursively solving $\frac{\partial Q(\theta, \theta^{I})}{\partial \theta} = 0$ for the model parameter $\theta^{I+1}$. Each iteration is guaranteed to improve log-likelihood, and the algorithm is guaranteed to converge to a local maximum.
%

\ifCLASSOPTIONcompsoc
  \section*{Acknowledgments}
\else
  \section*{Acknowledgment}
\fi

This research was supported by the U. S. Army Research
Office under grant W911NF-19-1-0365, and the National Science Foundation under grant 1714180.

\ifCLASSOPTIONcaptionsoff
  \newpage
\fi

\end{document}